# Gauge Physics of Spin Hall Effect

Seng Ghee Tan,[1,2†] Mansoor B. A. Jalil,[2,3] Congson Ho,[2] Zhuobin Siu,[2] Shuichi Murakami[4]

*(1) Data Storage Institute, A*STAR (Agency for Science, Technology and Research), DSI Building, 5 Engineering Drive 1, Singapore 117608*

*(2) Computational Nanoelectronics and Nano-device Laboratory, Electrical and Computer Engineering Department, National University of Singapore, 4 Engineering Drive 3,Singapore 117576*

*(3) Information Storage Materials Laboratory, Electrical and Computer Engineering Department, National University of Singapore, 4 Engineering Drive 3, Singapore 117576*

*(4) Department of Physics, Tokyo Institute of Technology, 2-12-1 Ookayama, Meguro-ku, Tokyo 152-8551, Japan*

Abstract

Spin Hall effect (SHE) has been discussed in the context of Kubo formulation, geometric physics, spin orbit force, and numerous semi-classical treatments. It can be confusing if the different pictures have partial or overlapping claims of contribution to the SHE. In this article, we present a gauge-theoretic, time-momentum elucidation, which provides a general SHE equation of motion, that unifies under one theoretical framework, all contributions of SHE conductivity due to the kinetic, the spin orbit force (Yang-Mills), and the geometric (Murakami-Fujita) effects. Our work puts right an ambiguity surrounding previously partial treatments involving the Kubo, semiclassical, Berry curvatures, or the spin orbit force. The full treatment shows the Rashba 2DEG SHE conductivity to be $+\frac{e}{8\pi}$ instead of $-\frac{e}{8\pi}$, and Rashba heavy hole $+\frac{9e}{8\pi}$ instead of $-\frac{9e}{8\pi}$.

† Corresponding author:
Seng Ghee Tan
Email: Tan_Seng_Ghee@dsi.a-star.edu.sg

## Spin Hall Effect (SHE)

Spin Hall Effect (SHE) [1-5] refers generally to the transverse separation of the electron carriers of opposite spin, quantized along the axis-z, which results in a net accumulation of spin but not charge on the left and right lateral edges of a nanoscale device. There have been many studies of the numerous possible mechanisms that could have given rise to SHE, but the gauge theory approach by Murakami et al. [6] showed for the first time that in the Luttinger spin orbit coupling (SOC) system, SHE physics is related to the adiabatic alignment of electron spin with the spin orbit effective magnetic field in the momentum space. An emergent form of magnetic field, with spin quantization axis along the lab-z axis, can then be defined and linked physically to a transverse velocity component of geometric origin. Following this emergent gauge approach, SHE physics of k-geometric origin could be conveniently extended to many other systems, e.g. the linear and the cubic spin orbit in semiconductor and metal, pseudospin in massless and massive graphene, topological insulator and so forth [7, 8].

On the other hand, Sinova et al. [9] derived the SHE conductivity for a two-dimensional-electron-gas (2DEG) system with linear Rashba SOC. Careful analysis [8, 10-13] would reveal that the SHE conductivity is in fact related to the velocity of kinetic origin. In 2010, Fujita et al. derived a gauge field in time (t) space that also led specifically to the kinetic velocity contributing to SHE in the 2DEG. The time-space gauge field can, in turn be linked to a t-geometric velocity which has the same form as [13, 14] the k-geometric velocity of Murakami. It is thus clear that one now should be particularly mindful of the multiple sources of velocity that contribute to the physics of SHE: kinetic, Murakami k-geometric, and Fujita t-geometric.

On the other hand, a separate body of work [15-19] which study the spin transverse force in terms of the non-Abelian spin orbit gauge, has led to the concepts of spin orbit force and spin orbit velocity. At first glance, one might be tempted to ascribe the transverse spin orbit force to SHE. But it was soon realized that while spin orbit force might contribute to the jittering motion (Zitterbewegung) of the spin carrier, it did not quite contribute to SHE yet. In fact, it is the spin orbit velocity that provides an additional source to the SHE. This results immediately in a SHE velocity originating from an emergent gauge reminiscent of the non-Abelian Yang-Mills gauge.

We are therefore motivated to provide, in this paper, a gauge-theoretic energy framework that unifies SHE velocity of kinetic, Yang-Mills, k-geometric, and t-geometric origins for any SOC system under one equation of motion (EOM). One unified energy system that merges the two spaces of t and k is derived, debunking any previous suspicion of overlapping energy terms. The energy equation with a merged t-k identity is then used to derive the velocity equation-of-motion (EOM) for all SHE systems. Previous efforts [13, 14] unified Luttinger and Rashba SHE with respect to the adiabatic physics and the gauge fields, but still it remained that the Luttinger was described in k-space, and the Rashba in t-space.

Our work is thus divided into three sections. Section 2 is dedicated to unifying the locally gauge transformed energy landscape of a SOC system, providing the theoretical basis for a form-invariant, t-k manifestation of the gauge potential. Local transformation in this context is an abstract but useful technique to absorb the physics of spin dynamics into the gauge potential. Section 3 is dedicated to using the t-k from-invariant energy to derive the EOM for the SHE of the spin carrier. We show here that only a properly transformed, form-invariant t-k energy should be used to derive the SHE EOM that produces in clear-cut, and non-overlapping manner, the velocity components of kinetic, Yang-Mills, and geometric origins. Section 4 is to provide, using specifically the two-dimensional hetero-structure with Rashba SOC, a physical illustration of SHE resulting from all contributing velocity.

## Energy in the Unified Time-Momentum (t-k) Space

The Hamiltonian of a system with SOC can be written with the physical clarity of simple magnetism as follows:

$$H = \frac{\boldsymbol{p}^2}{2m} - \gamma\boldsymbol{\sigma}.\boldsymbol{B} + e\boldsymbol{E}^{\boldsymbol{a}}.\boldsymbol{r} \tag{2}$$

where $\boldsymbol{B}$ is a momentum dependent effective magnetic field, and $\gamma$ has the dimension of $[\gamma] = \frac{Joule}{Tesla}$. In a single-particle system with electric field, the Hamiltonian $H = \frac{\boldsymbol{p}^2}{2m} + e\boldsymbol{E}^{\boldsymbol{a}}.\boldsymbol{r}$ might seem sufficient, at first glance, to describe a carrier with kinetic and potential energy. But a charge-spin carrier with a constant *p* in the presence of *E* field generates an energy term of $\gamma\boldsymbol{\sigma}.\boldsymbol{B}(\boldsymbol{p})$. The Dirac relativistic quantum mechanics is needed to account for this SOC energy. In the absence of any retardation effect due to scattering, the carrier would accelerate due to the *E* field. As a result, the carrier will acquire an *E*-dependent energy $\gamma\boldsymbol{\sigma}.\boldsymbol{B}\left(\frac{d\boldsymbol{p}}{dt}\right)$. This is a geometric related energy that can only be revealed with the local gauge transformation or time-dependent perturbation treatment.

*Momentum Space*

The approach that has been used to derive SHE velocity in **Refs [6, 20]** is based on a local gauge transformation in the k-space. In the "Schrodinger" picture, transformation applies to the three-k space only, but not the t-space, because momentum is time-independent. Local transformation leads to

$$H_S' = UH_SU^\dagger = \frac{\boldsymbol{p}^2}{2m} - \gamma\sigma_z B + e\boldsymbol{E}_{\boldsymbol{a}}.(\boldsymbol{r} + iU\partial_{\boldsymbol{k}}U^\dagger) \tag{3}$$

where the gauge potential $\boldsymbol{\mathcal{A}}_{\boldsymbol{k}} = -iU\partial_{\boldsymbol{k}}U^\dagger$, with dimension $[\boldsymbol{\mathcal{A}}_{\boldsymbol{k}}] = m^{-1}$, is associated with the SHE velocity via the Karplus-Luttinger method.

*Time Space*

The t-space approach has on the other hand, been adopted in **Refs [10-13]** to derive SHE in **Refs [9, 21, 22]**. To study the transformation in the fourth time space, an "Interaction" picture is necessary. The term $\gamma\boldsymbol{\sigma}.\boldsymbol{B}(\boldsymbol{k})$ would become $\gamma\boldsymbol{\sigma}.\boldsymbol{B}(t)$, where $\boldsymbol{B}(t) = \left(\boldsymbol{B}(\boldsymbol{k}) - \frac{\mathrm{eE_\nu t}}{\hbar}.\frac{\partial \boldsymbol{B}(\boldsymbol{k})}{\partial \mathrm{k_\upsilon}}\right)$ is the instantaneous SOC field. One needs to split the Hamiltonian into two parts, i.e. $H_S = H_0 + V_S$, and note the following:

$$H_I(t) = H_0 + e^{iH_0 t} V_S e^{-iH_0 t} \tag{4}$$

or $H_I(t) = H_0 + V_I(t)$. In the "Interaction" picture, one has $i\hbar\partial_t\psi_I(t) = V_I(t)\psi_I(t)$. Thus a transformation in time space is appropriate here

$$i\hbar U\partial_t U^\dagger\, \psi_I'(t) = V_I'(t)\, \psi_I'(t) \tag{5}$$

where $i\hbar U\partial_t U^\dagger = i\hbar\partial_t + i\hbar U\left(\partial_t U^\dagger\right)$, and the second term is contingent upon $\partial\boldsymbol{k}/\partial t \neq 0$, a condition that would be fulfilled when $E$ field is present in the device. On the RHS of Eq.(5), $V_I'(t) = V_I(t) + i\hbar U(\partial_t U^\dagger)$, where $i\hbar U(\partial_t U^\dagger)$ is also known as the time-space gauge with the dimension of energy. The unitary rotation operator $U$ in local space is

$$U = exp\frac{i}{2}(\boldsymbol{\sigma}.\boldsymbol{\omega})t \tag{6}$$

and one can quickly obtain that $i\hbar U(\partial_t U^\dagger) = \frac{1}{2}\hbar U\boldsymbol{\sigma}U^\dagger.\boldsymbol{\omega}$. Rearranging, one can now write the locally time-transformed Hamiltonian in "Interaction" picture as

$$H_I(t) = \left(\frac{\boldsymbol{p}_{\boldsymbol{I}}^{\boldsymbol{2}}}{2m} - \gamma\sigma_z B(t)\right) - i\hbar U(\partial_t U^\dagger) + e\boldsymbol{E}^{\boldsymbol{a}}.\boldsymbol{r} \tag{7}$$

Equation (7) in time space has the same form as Equation (3) in momentum space. Re-examining the time-gauge in the k-space,

$$-i\hbar(U^\dagger\partial_t U) = -i\frac{\partial\hbar\boldsymbol{k}}{\partial t}.U\partial_{\boldsymbol{k}}U^\dagger = e\boldsymbol{E}^{\boldsymbol{a}}.iU(\partial_{\boldsymbol{k}}U^\dagger) \tag{8}$$

leads to one obtaining a gauge expression in the k-space. Note that $\left(-\frac{\partial\hbar\boldsymbol{k}}{\partial t}\right)=e\boldsymbol{E}$. The fact that $H_0=e\boldsymbol{E^a}.\boldsymbol{r}$ is important for the survival of the gauge $-i\frac{\partial\hbar\boldsymbol{k}}{\partial t}.U\partial_{\boldsymbol{k}}U^{\dagger}$ for the gauge coupling constant is only non-vanishing because of $H_0=e\boldsymbol{E^a}.\boldsymbol{r}$. Inverse transformation of $H'_S=e^{-iH_0t}H_I(t)e^{iH_0t}$ would lead to higher order terms with respect to $-i\hbar U(\partial_t U^{\dagger})$. Dropping the higher order terms, the following is arrived

$$H'_S=\frac{\boldsymbol{p^2}}{2m}-\gamma\sigma_z B+eE^a_k.(r^k+iU\partial_k U^{\dagger}) \tag{9}$$

Note that for expression $iU\partial_k U^{\dagger}$, subscripts are subject to $k\in(k_x,k_y,k_z)$, while for expression $r^k$, subscripts are subject to $k\in(x,y,z)$. Equation (9) is transformed in the t-space at the outset, but now appears identical to the k-space Equation (3). The energy equations have thus been merged under a form-invariant t-k identity. What is clear from the above is that the gauge potential derived, interchangeably in the k-space or the t-space, will not exist simultaneously in both spaces. It is now logical to conclude that the gauge potential has an independent, non-overlapping contribution to the geometric velocity of SHE. Back to the Lab frame (mere relabeling of the axis), one can write the Hamiltonian in the form of Zeeman magnetic field as in Eq.10(a), and in the form of Lorentz magnetic field as in Eq.10(b),

$$H=\frac{\boldsymbol{p^2}}{2m}-\boldsymbol{\sigma}.\left(\boldsymbol{B}+\frac{e}{\gamma}E^a_k\boldsymbol{A_k}\right)+e\boldsymbol{E^a}.\boldsymbol{r} \tag{10a}$$

$$H=\frac{\boldsymbol{p^2}}{2m}-\gamma\boldsymbol{\sigma}.\boldsymbol{B}+eE^a_k(r^k-\boldsymbol{\sigma}.\boldsymbol{A}_k) \tag{10b}$$

Note that in the above the convention $E^a_k(iU^{\dagger}\partial_k U)=-eE^a_k\boldsymbol{\sigma}.\boldsymbol{A}_k$ is followed. Similar convention in time space is followed $\hbar(iU^{\dagger}\partial_t U)=-\hbar\boldsymbol{\sigma}.\boldsymbol{A}_t$. Table 1. provides a summary of the theoretical processes for both spaces.

*Table 1. Summary of important gauge theoretic quantities expressed in both time and momentum spaces.*

.

| | | **Time-space** | **Momentum-space** |
|---|---|---|---|
| 1. | Local gauge transformation and vector potential notation | $\hbar(iU^{\dagger}\partial_t U) = -\hbar\mathcal{A}'_t$ | $eE_k^a(iU^{\dagger}\partial_k U) = -eE_k^a\mathcal{A}'_k$ |
| 2. | Physics of effective magnetic field $\mathcal{A}'_t = \boldsymbol{\sigma}.\boldsymbol{A}'_t$ , $eE_k^a\mathcal{A}'_k = E_k^a(\boldsymbol{\sigma}.\boldsymbol{A}'_k)$ | $\frac{\hbar}{2}(iU\boldsymbol{\sigma}U^{\dagger}).(\boldsymbol{n}\times\partial_t\boldsymbol{n}) = -\hbar\boldsymbol{\sigma}.\boldsymbol{A}'_t$ | $eE_k^a\frac{1}{2}(iU\boldsymbol{\sigma}U^{\dagger}).(\boldsymbol{n}\times\partial_k\boldsymbol{n}) = -eE_k^a\boldsymbol{\sigma}.\boldsymbol{A}'_k$ |
| 3. | Hamiltonian in the locally rotated frame | $H = \frac{\boldsymbol{p}^2}{2m} - \gamma\sigma_z B + e\boldsymbol{E}^{\boldsymbol{a}}.\boldsymbol{r} + \hbar\mathcal{A}'_t$ <br> Rotation of the $z$ −axis to $B(t)$ | $H = \frac{\boldsymbol{p}^2}{2m} - \gamma\sigma_z B + e\boldsymbol{E}^{\boldsymbol{a}}.\boldsymbol{r} - eE_k^a\mathcal{A}'_k$ <br> Rotation of the $z$ −axis to $B(k)$ |
| 4. | Hamiltonian in lab frame, showing effective magnetic fields | $H = \frac{\boldsymbol{p}^2}{2m} - \gamma\boldsymbol{\sigma}.\boldsymbol{B} + e\boldsymbol{E}^{\boldsymbol{a}}.\boldsymbol{r} + \hbar\boldsymbol{\sigma}.\boldsymbol{A}_t$ | $H = \frac{\boldsymbol{p}^2}{2m} - \gamma\boldsymbol{\sigma}.\boldsymbol{B} + e\boldsymbol{E}^{\boldsymbol{a}}.\boldsymbol{r} - eE_k^a(\boldsymbol{\sigma}.\boldsymbol{A}_k)$ <br> $\boldsymbol{k}$ in three-momentum space |

## Spin Hall Equation of Motion

In the energy physics of the SOC systems, we have shown in Section 2 the unified, t-k manifestation of the local gauge. It is reasonable to ascribe before merger, the k-gauge to the k-geometric velocity due to Murakami [6], and the t-gauge to the t-geometric velocity due to Fujita [10]. What we have done, after merger is uniting the two velocities and precluding their simultaneous manifestation. We will show in this section that a physically intuitive SHE velocity EOM that encompasses the kinetic, Yang-Mills, and Murakami-Fujita velocity can be derived from the locally transformed t-k energy equation. One is free to view the effective Hamiltonian in either time or momentum space. In time space, one can define an effective magnetic field of $\boldsymbol{B}_{\boldsymbol{\Sigma}} = \boldsymbol{B} - \frac{\hbar}{\gamma}\boldsymbol{A}_{\boldsymbol{t}}$ . The complete SHE velocity EOM is

$$\langle v_y^z\rangle = \frac{1}{2}\left\langle\Sigma\left|\left\{\frac{p_y}{m},\, \sigma_z\right\}\right|\Sigma\right\rangle - \frac{1}{2}\langle\Sigma|\,\{\gamma\,(\partial B_\Sigma^a/\partial p_y)\,\sigma_a\,,\sigma_z\}\,|\Sigma\rangle$$

(11)

$$\langle v_y^z\rangle \;\equiv\; \frac{p_y}{m}\langle\Sigma|\,\sigma_z|\Sigma\rangle - \gamma\frac{\partial B_\Sigma^z}{\partial p_y}$$

(12)

The first term on the RHS is the kinetic velocity. The second term comprises the Yang-Mills and the Murakami-Fujita velocity as shown below

$$\langle v_z^y\rangle = \langle v_z^y\rangle_{KE} + \langle v_z^y\rangle_{YM} + \langle v_z^y\rangle_{MF}$$

(13)

$kinetic$ ↙ $\quad$ $Yang - Mills$ ↓ $\quad$ $Murakami - Fujita$ ↘

The kinetic SHE velocity can also be written as

$$\langle v_y^z\rangle_{KE} = \frac{p_y}{m}\langle\pm|\,\boldsymbol{\sigma}.\boldsymbol{a_z}\,|\pm\rangle = \frac{p_y}{m}(\boldsymbol{n}_\Sigma.\boldsymbol{a_z})$$

(14)

with both $\boldsymbol{n}_\Sigma$ and $\boldsymbol{a_z}$ having unity magnitude. The $\pm$ sign arising from the two eigenstates of $|\Sigma\rangle$ correspond to the $(+)$ and the $(-)$ bands, respectively. The negative sign of the spin orbit energy **i**mplies that energy is low when spin is aligned with the effective *B* field. Thus in the $(+)$ band of Fig.1a, spin is aligned along the *B* fields in both East and West. In the $(-)$ band of Fig.1b, spin is anti-aligned everywhere. The term $\boldsymbol{n}_\Sigma(p_y)\,.\,\boldsymbol{a_z}$ is the fractional unit of the *B* field projected to the $\boldsymbol{a_z}$. Note that $\boldsymbol{B}_\Sigma = \frac{\hbar}{\gamma}(\boldsymbol{n}\times\partial_t\boldsymbol{n}) + B\boldsymbol{n}$, and $\boldsymbol{n}_\Sigma.\boldsymbol{a_z} = \frac{B_\Sigma^z}{B_\Sigma}$, lead to $\boldsymbol{n}_\Sigma = \frac{\hbar}{\gamma B_\Sigma}(\boldsymbol{n}\times\partial_t\boldsymbol{n}) + \frac{B}{B_\Sigma}\boldsymbol{n}$, where $\frac{\hbar}{\gamma B_\Sigma}(\boldsymbol{n}\times\partial_t\boldsymbol{n})$ is the time-gauged spin orbit field, and $\frac{B}{B_\Sigma}\boldsymbol{n}$ is the simple spin orbit field. The kinetic spin velocity is thus

$$\langle v_y^z\rangle_{KE} = \pm\frac{p_y}{m}\left(\frac{\hbar}{\gamma B_\Sigma}\left(\boldsymbol{n}\times\frac{\partial\boldsymbol{n}}{\partial k_x}\right)E_x + \frac{B}{B_\Sigma}\boldsymbol{n}\right).\boldsymbol{a_z}$$

(15)

clearly showing that the external electric field is required to generate $\partial_t\boldsymbol{n}$. The expression $\boldsymbol{n}_\Sigma = \frac{\hbar}{\gamma B_\Sigma}(\boldsymbol{n}\times\partial_t\boldsymbol{n})$ relates to the physics of $E_x$ producing an effective $B$ field. For illustration in the 2D projected region surrounded by the equator, $\boldsymbol{n}_\Sigma$ points along axis -z in the region of $+p_y$, and axis +z in the region of $-p_y$ (see illustration in Fig.1). In other words, $\boldsymbol{n}_\Sigma$ actually changes sign with $p_y$. Therefore, careful examination of the $(+)$ band (Fig.1a) would show that $p_y\boldsymbol{n}_\Sigma.\boldsymbol{a_z}$, hence $\langle v_y^z\rangle_{KE}$ is negative in both the East and West hemisphere. The result is always positive in the $(-)$ band. To determine SHE conductance, one needs to sum $\langle v_y^z\rangle_{KE}$ over the entire Fermi surface for both bands. Since the two band cancels one another, it is necessary to identify a region where only one band exists, or in other words, to impose a band filtering effect.

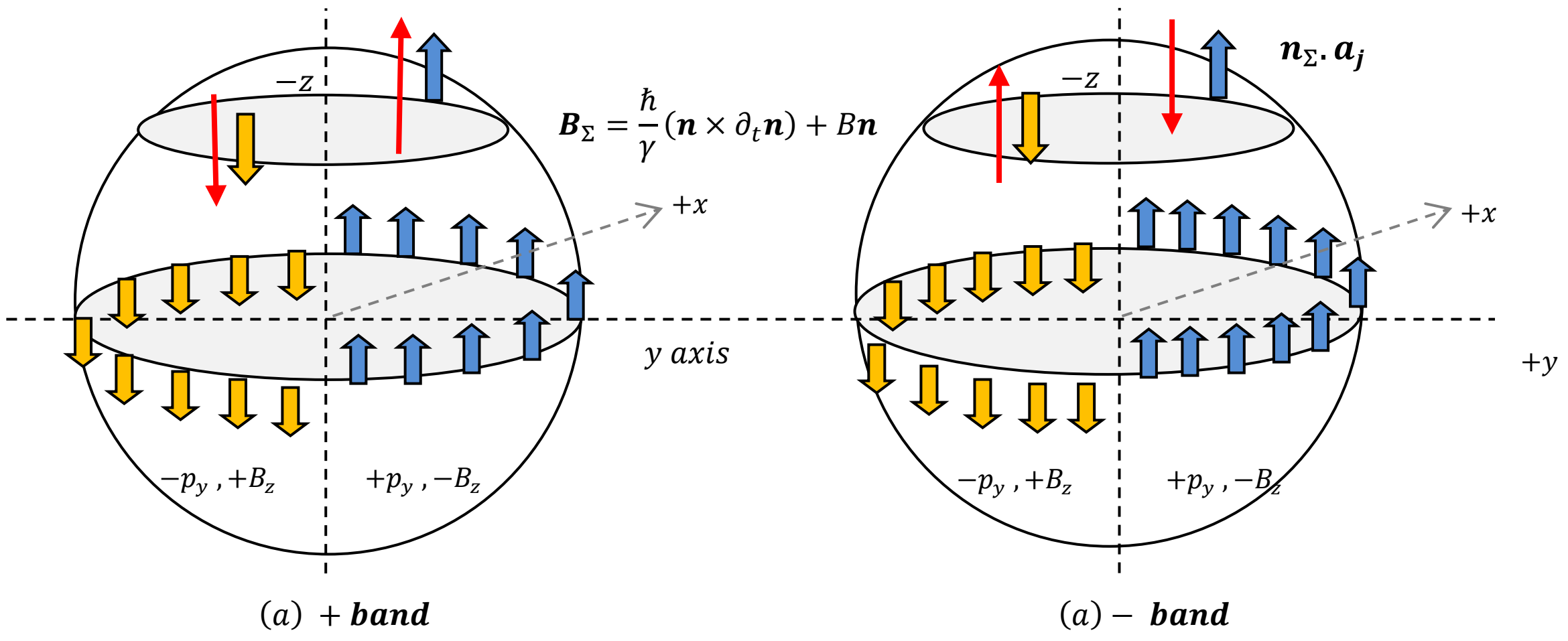

*Fig.1. Fermi sphere of a general electron gas system in the presence of spin orbit coupling shows a distribution of the momentum, band, and effective magnetic field projected along z ($B_z$). It is assumed that $p = \hbar k$. The shaded region encircled by the equator shows a specific system (Rashba 2D) where for (a) the $+$ band, $B_z$ changes sign over the Eastern and Western hemisphere, resulting in a positive kinetic spin velocity, (b) the $-$ band, $B_z$ changes sign in a similar manner, thus resulting in a negative kinetic spin velocity. The slender red arrow indicates spin polarization of $\pm\boldsymbol{n}_\Sigma.\boldsymbol{a}_j$.*

The kinetic velocity has been identified in previous time-space **[10]** to contribute to SHE in the following systems, producing SHE conductivity of $\sigma_{xy}^z = \frac{-e}{8\pi}$ **[9]** in 2D RSOC hetero-structure, $\sigma_{xy}^z = \frac{-ek_F}{12\pi^2}$ **[21]** in n-doped cubic-Dresselhaus, and $\sigma_{xy}^z = \frac{-9e}{8\pi}$ **[22]** in Rashba heavy-hole system. There was, however, no explicit previous effort to investigate the total SHE effects that should include the Yang-Mills and the Murakami-Fujita contribution in those systems. We will, in section 4, make explicit the other contributions to SHE in the 2D Rashba SOC system, now keenly studied for technological applications in magnetic memory. The SHE velocity of Yang-Mills $\langle v_y^z \rangle_{YM}$ and Murakami-Fujita $\langle v_y^z \rangle_{MF}$ inherit their negative signs from the spin orbit energy. The explicit expression is

$$\langle v_y^z \rangle = -\frac{e}{\hbar}\left(\underbrace{\left(\frac{\partial \boldsymbol{n}}{\partial k_y}\times\frac{\partial \boldsymbol{n}}{\partial k_x}\right)E_x+\left(\boldsymbol{n}\times\frac{\partial^2 \boldsymbol{n}}{\partial k_y \partial k_x}\right)E_x}_{\text{Murakami-Fujita}}+\underbrace{\frac{\gamma\partial B\boldsymbol{n}}{\partial k_y}}_{\text{Yang-Mills}}\right).\boldsymbol{a_z} \qquad (16)$$

The $\langle v_y^z \rangle_{MF}$ in the t-k form-invariant EOM has previously been studied in the k-space for the SOC systems of Luttinger [6], Perel-Dresselhaus [7, 8], cubic-Dresselhaus in n-doped Zinc-Blende [20]. There was, however, no previous effort that explicitly investigates the effect of kinetic contribution in those systems. On the other hand, the $\langle v_y^z \rangle_{YM}$ shown previously in simple (not locally gauged) quantum mechanics either vanishes for linear SOC systems, or is simply not considered. The reason for partial treatment in previous works could be due to conceptual ambiguity. It was not clear beforehand if the $\langle v_y^z \rangle_{KE}$ derived in the t-space gauge or the Kubo approach and the $\langle v_y^z \rangle_{MF}$ derived in the k-space gauge are strictly independent without overlapping contribution. This becomes clear after the form-invariant t-k energy unifies both spaces. The SHE velocity EOM descended from the t-k energy shows vividly that these velocities are additive. The simple conclusions one draws here is that previous SHE conductivity related to Kubo, semiclassical, t-gauge are mostly related to $\langle v_y^z \rangle_{KE}$. On the other hand, previous SHE conductivity related to Berry curvature or k-gauge are related to $\langle v_y^z \rangle_{MF}$. What we would like to establish therefore in this work is that future treatment of SHE should be based on a locally transformed, t-k Hamiltonian that would provide all velocity $(\langle v_y^z \rangle_{KE} + \langle v_y^z \rangle_{YM} + \langle v_y^z \rangle_{MF})$ contributing to the physics of SHE.

## SHE in the 2D Rashba system

One popular form of SHE exists in device or hetero-structure that exhibits the Rashba spin orbit coupling (RSOC). Examples are the GaAs/AlGaAs/GaAs semiconductor heterostructure, or the oxide/Metal/Pt metal multilayer, both with structural inversion asymmetry. Shown below is a schematic of the nanoscale structure where the specific RSOC exists as a result of inversion asymmetry at the interface. The effective magnetic field of the RSOC device is

$$\boldsymbol{B}_\Sigma = \left(-\frac{\alpha k_y}{\gamma}\boldsymbol{a_x} + \frac{\alpha k_x}{\gamma}\boldsymbol{a_y}\right) + \frac{eE_x k_y}{\gamma k^2}\boldsymbol{a_z} \tag{17}$$

where $\alpha$ is the strength of the Rashba SOC effect.

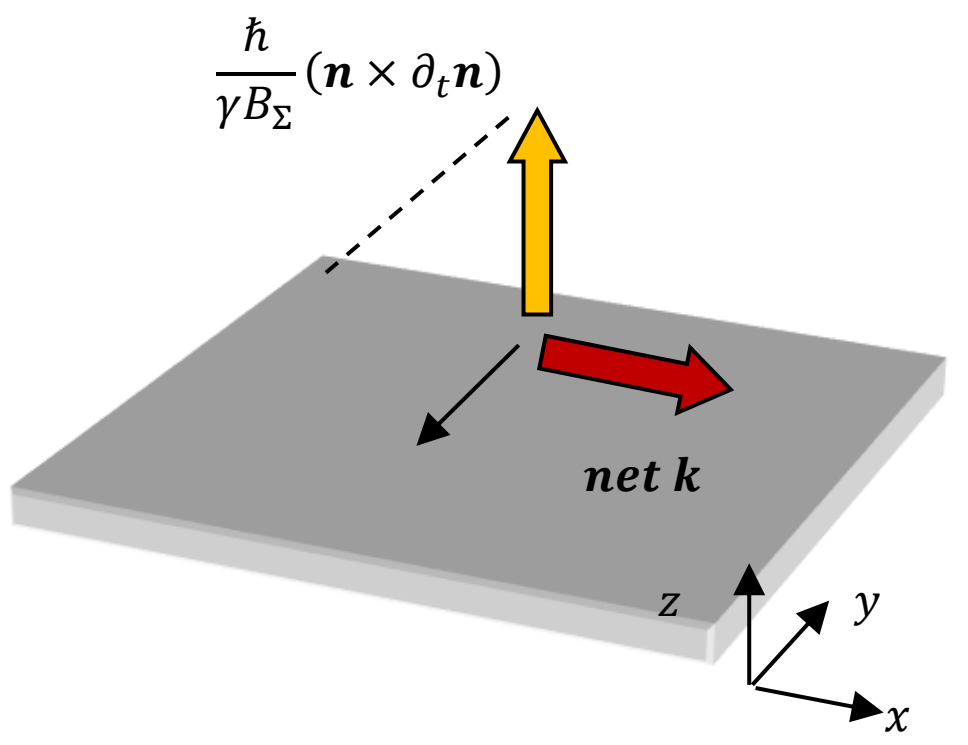

(a)

| | Quantities | Explicit |
|---|---|---|
| 1. | $\boldsymbol{n} = \frac{\gamma}{\alpha k}\boldsymbol{B}$ | $\left(\frac{-k_y}{k}, \frac{k_x}{k}, 0\right)$ |
| 2. | $\frac{\partial \boldsymbol{n}}{\partial k_x}$ | $\left(\frac{k_x k_y}{k^3}, \frac{k_y^2}{k^3}, 0\right)$ |
| 3. | $\frac{\partial \boldsymbol{n}}{\partial k_y}$ | $\left(\frac{-k_x^2}{k^3}, \frac{-k_x k_y}{k^3}, 0\right)$ |

(b)

Fig.2. (a) In the case of a 2D nanostructure, where $n$ lies in the $x - y$ plane, the $(n \times \partial_t n)$ term points along $z$, thus one has $\langle v_y^z \rangle_{KE} = \pm \frac{\hbar p_y}{\gamma B}\ |n \times \partial_t n|$; (b) Table of quantities for the Rashba system that can be used to derive the SHE expression for the Rashba system.

We will first examine a two-dimensional system, where the SHE current density is obtained from the SHE kinetic velocity as follows

$$J_y^z = \int g \frac{dk_x dk_y}{(2\pi)^2}\ \langle v_y^z \rangle_{KE} \tag{18}$$

The coupling constant $g$ represents spin flux for $g = \hbar/2$ , or electron charge flux for $g = e$. It has been shown that, for $g = \hbar/2$, SHE conductivity $\sigma_y^z = -\frac{e}{8\pi}$ is resulted in the annular region of the Rashba bandstructure where only the $(+)$ band exists below the Fermi energy. In the region where both $(+)$ and $(-)$ are below the Fermi energy, total SHE conductivity vanishes. In fact, the above has the dimension of $\frac{[g]}{[L]\,[t]}$, and the general SHE conductivity is $\sigma_y^z = -\frac{ge}{2h}$, leading to respectively, the charge $(g = e)$ and the spin $(g = \hbar/2)$ flux of:

$$\sigma_y^z = -\frac{1}{2}\frac{e^2}{h} \quad , \quad and \quad \sigma_y^z = -\frac{e}{8\pi} \tag{19}$$

Referring to Eq.(15), we will now proceed to the Yang-Mills velocity. In the case of a linear SOC system, where effective SOC field is contained in the 2D plane, it is easy to determine that the Yang-Mills effect vanishes. We will move on to the last SHE term which is the Murakami-Fujita of $\langle v_y^z \rangle_{MF} = -\frac{e}{\hbar}\left(\left(\frac{\partial \boldsymbol{n}}{\partial k_y} \times \frac{\partial \boldsymbol{n}}{\partial k_x}\right) E_x + \left(\boldsymbol{n} \times \frac{\partial^2 \boldsymbol{n}}{\partial k_y \partial k_x}\right) E_x\right) . \boldsymbol{a_z}$, and which can in turn be broken down into two terms. The first term produces SHE current density

$$J_y^z = \int -g \frac{dk_x dk_y}{(2\pi)^2} \frac{e}{\hbar}\left(\frac{\partial \boldsymbol{n}}{\partial k_y} \times \frac{\partial \boldsymbol{n}}{\partial k_x}\right) E_x . \boldsymbol{a_z} \tag{20}$$

It can be shown that $\left(\frac{\partial \boldsymbol{n}}{\partial k_y} \times \frac{\partial \boldsymbol{n}}{\partial k_x}\right) . \boldsymbol{a_z} = -\delta(k_x)\delta(k_y)\,\pi$ . Considering that there are two bands cutting through the k=0 point, the SHE conductivity due to $\langle v_y^z \rangle_{MF}$ is $+\frac{e}{4\pi}$. This is in addition to the $-\frac{e}{8\pi}$ arising due to $\langle v_y^z \rangle_{KE}$ giving rise to a total SHE $\sigma_y^z = +\frac{e}{8\pi}$. The advantage of physical clarity with the gauge theoretic approach is clearly manifest here. The first contribution to SHE conductivity originates from the kinetic velocity which is effective in the annular region of the 2D concentric circles. The second contribution originates from the Murakami-Fujita velocity that has a geometric origin, and is effective in the degenerate point where the $(+)$ and the $(-)$ bands intersect. The same is carried out for the Rashba heavy hole system where $\left(\frac{\partial \boldsymbol{n}}{\partial k_y} \times \frac{\partial \boldsymbol{n}}{\partial k_x}\right) . \boldsymbol{a_z} = -3\delta(k_x)\delta(k_y)\,\pi$ . The final results are SHE $\sigma_y^z = +\frac{9e}{8\pi}$ instead of the $-\frac{9e}{8\pi}$ with partial treatments. The gauge theoretic method provides a full treatment of the SHE, showing results opposite in sign to previous partial treatments. We will now proceed to the second part of $\langle v_y^z \rangle_{MF}$, which is

$$J_y^z = \int -g\,\frac{dk_x dk_y}{(2\pi)^2}\,\frac{e}{\hbar}\left(\boldsymbol{n}\times\frac{\partial^2\boldsymbol{n}}{\partial k_i \partial k_x}\right)E_x.\boldsymbol{a_z} = \int -g\,\frac{dk_x dk_y}{(2\pi)^2}\,\frac{eE_x}{\hbar}\left(\frac{k_y^2 - k_x^2}{k^4}\right) \tag{21}$$

When integrated via partial fractional reformulation of the integrand, the integral delivers a " $\pi$" or a "$-\pi$" solution depending on the sequence in which the integration is performed. But a proper treatment referring to the Fubini-Tonelli theorem leads to its vanishing results.

## CONCLUSION

The central results in this paper consist of the unified energy and the SHE EOM. On the energy, we have provided a theoretical basis to the existence of the t-k interchangeable, form-invariant Hamiltonian. This Hamiltonian allows the physics of SHE in various SOC systems to be studied under one SHE velocity EOM. The EOM descended from the t-k energy would lead to the kinetic, Yang-Mills, and Murakami-Fujita velocity which give a complete account of all contribution to SHE conductivity in any SOC system. Our work puts right an ambiguity surrounding previously partial treatments of SHE involving the use of the Kubo, semiclassical, Berry curvatures, or the spin orbit gauge. We showed for the Rashba 2DEG and the Rashba heavy hole that full treatment produces SHE conductivity of opposite signs due to the Murakami-Fujita contribution.

## Acknowledgement

We would like to thank Takashi Fujita for comment on gauge field in different spaces.